\documentclass[aps,prb,a4paper,10pt,twocolumn,showpacs,floatfix,superscriptaddress,preprintnumbers,longbibliography]{revtex4-2}
\usepackage{float}
\usepackage{dcolumn,graphicx,color,booktabs,microtype,afterpage}
\usepackage{amssymb}
\usepackage{amsmath}
\usepackage[charter,greekuppercase=italicized]{mathdesign}
\usepackage{sidecap}
\usepackage[mathlines]{lineno}

\graphicspath{{./}{figure/}}
\renewcommand{\tablename}{Table}
\makeatletter\renewcommand{\fnum@figure}[1]{\figurename~\thefigure.~}\makeatother
\makeatletter\renewcommand{\fnum@table}[1]{\tablename~\thetable.}\makeatother

\newcount\hh \newcount\mm
\hh=\time \divide\hh by 60
\mm=\hh \multiply\mm by 60 \mm=-\mm
\advance\mm by \time
\def\now{\number\hh:\ifnum\mm<10{}0\fi\number\mm}

\usepackage[colorlinks,plainpages=false,linkcolor=blue,urlcolor=blue,citecolor=blue,pdfpagemode=UseNone,pdfstartview=FitBH]{hyperref}

\newcommand{\tcr}[1]{\textcolor{black}{#1}}

\begin{document}

\makeatletter\renewcommand{\ps@plain}{%
\def\@evenhead{\hfill\itshape\rightmark}%
\def\@oddhead{\itshape\leftmark\hfill}%
\renewcommand{\@evenfoot}{\hfill\small{--~\thepage~--}\hfill}%
\renewcommand{\@oddfoot}{\hfill\small{--~\thepage~--}\hfill}%
}\makeatother\pagestyle{plain}

\preprint{\textit{Preprint: \today, \now}} 
\title{Time-reversal symmetry breaking in Re-based superconductors: Recent developments}
%
\author{T.\ Shang}\email[Corresponding authors:\\]{tshang@phy.ecnu.edu.cn}
\affiliation{Key Laboratory of Polar Materials and Devices (MOE), School of Physics and Electronic Science, East China Normal University, Shanghai 200241, China}
\affiliation{Laboratory for Multiscale Materials Experiments, Paul Scherrer Institut, Villigen CH-5232, Switzerland}
\author{T.\ Shiroka}\email[Corresponding authors:\\]{tshiroka@phys.ethz.ch}
\affiliation{Laboratory for Muon-Spin Spectroscopy, Paul Scherrer Institut, Villigen PSI, Switzerland}
\affiliation{Laboratorium f\"ur Festk\"orperphysik, ETH Z\"urich, CH-8093 Z\"urich, Switzerland}
\begin{abstract}
In the recent search for unconventional- and topological superconductivity, 
noncentrosymmetric superconductors (NCSCs) rank among the most promising 
candidate materials. Surprisingly, some of them --- especially those 
containing rhenium 
--- seem to exhibit also time-reversal symmetry (TRS) 
breaking in their superconducting state, while TRS is preserved in many 
other isostructural NCSCs. To date, a satisfactory explanation for such 
discrepant behavior, albeit crucial for understanding the unconventional 
superconductivity of these materials, is still missing.
Here we review the most recent developments regarding the Re-based class, 
where the muon-spin relaxation ($\mu$SR) technique plays a key 
role due to its high sensitivity to the weak internal fields 
associated with the TRS breaking phenomenon. 
We discuss different cases of Re-containing superconductors, comprising 
both centrosymmetric- and noncentrosymmetric crystal structures and 
ranging from pure rhenium, to Re$T$ ($T$ = 3$d$-5$d$ early transition metals), 
to the dilute-Re case of ReBe$_{22}$. $\mu$SR results suggest that the 
rhenium presence and its amount are two key factors for the appearance 
and the extent of TRS breaking in Re-based superconductors.
Besides summarizing the existing findings, we also put forward 
future research ideas regarding the exciting field of materials showing TRS breaking.
\end{abstract}

\maketitle\enlargethispage{3pt}

\vspace{-5pt}
\section{\label{sec:Introduction}Introduction}\enlargethispage{8pt}
%

The combination of intriguing fundamental physics with far-reaching 
potential applications has made unconventional superconductors one of 
the most studied classes of materials. Standing out among them are the 
noncentrosymmetric superconductors (NCSCs)~\cite{Bauer2012}, whose 
crystal structures lack an inversion symmetry. As a consequence, in NCSCs, the 
strict symmetry-imposed requirements are relaxed, allowing 
mixtures of spin-singlet and spin-triplet Copper pairing channels, 
thus setting the scene for a variety of exotic properties, as e.g., 
upper critical fields beyond the Pauli limit, nodes in the superconducting 
gaps, etc. (see Refs.~\cite{Bauer2012,Smidman2017,Ghosh2020} for an overview).
The degree of mixing in such combined pairings is related to the 
strength of the antisymmetric spin-orbit coupling (ASOC) and to other 
microscopic parameters, still under investigation. Currently, 
NCSCs rank among the foremost categories of superconducting materials 
in which to look for topological superconductivity (SC) or to realize the
Majorana fermions, pairs of the latter potentially acting as 
noise-resilient qubits in quantum computing~\cite{Kim2018,Sun2015,Ali2014,Sato2009,Tanaka2010,Sato2017,Qi2011,Kallin2016}.

In general, the various types of NCSCs can be classified into two classes. 
One consists of strongly correlated materials, as e.g., CePt$_3$Si~\cite{Bauer2004}, or
Ce(Rh,Ir)Si$_3$~\cite{Muro1998}, 
which belong to the heavy-fermion compounds. Owing to the strong 
correlation and the interplay between $d$- and $f$-electrons, these materials 
often exhibit rich magnetic and superconducting properties. Since 
their superconductivity is most likely mediated by 
spin fluctuations, this implies an unconventional (i.e., non phonon-related) pairing mechanism. 
Conversely, the other class consists mainly of weakly correlated materials, 
which are free of ``magnetic'' $f$-electrons, as e.g., LaNiC$_2$, La$_7$Ir$_3$, 
CaPtAs, or Re$T$ ($T$ = 3$d$-5$d$ early transition metals)~\cite{Hillier2009,Barker2015,Shang2020,Singh2014,Singh2017,Shang2018a,Shang2018b}. 
Obviously, their superconductivity
is not mediated by electrons' spin fluctuations.  
Hence, they lend themselves as prototype parent systems where one can study the intrinsic pairing mechanisms in NCSCs.

Recently, superconductivity with broken time-reversal symmetry (TRS) 
has become a hot topic in NCSCs. The main reason for this is the 
discovery of TRS breaking in some
weakly-correlated NCSCs using 
muon-spin relaxation ($\mu$SR). 
Surprisingly,
the superconducting properties of the latter
largely resemble those of conventional superconductors, i.e., they 
are clearly distinct from those of the above mentioned 
strongly-correlated NCSCs. To date, only a handful of
NCSC families have been shown to exhibit TRS breaking in the superconducting state, including LaNiC$_2$~\cite{Hillier2009}, La$_7$(Rh,Ir)$_3$~\cite{Barker2015,Singh2018La7Rh3}, Zr$_3$Ir~\cite{Shang2020b}, CaPtAs~\cite{Shang2020}, and Re$T$~\cite{Hillier2009,Barker2015,Shang2020,Singh2014,Singh2017,Shang2018a,Shang2018b}. Except for the recently studied CaPtAs, where coexisting TRS breaking and superconducting gap nodes were observed below $T_c$, in most of the above cases the 
superconducting properties evidence a conventional $s$-wave 
pairing, characterized by a fully opened superconducting gap. 
This leads to an interesting fundamental question: does the observed 
TRS breaking in NCSCs originate from an unconventional superconducting mechanism  
(i.e., from a pairing other than that mediated by phonons), 
or it can occur also in presence of conventional pairing 
(via some not yet understood mechanism)~\cite{Smidman2017,Ghosh2020}?
Why, among the many different NCSCs families, only a few exhibit 
a broken TRS in the superconducting state, also remains an intriguing 
open question.

In general, the causes behind the
TRS breaking at the onset of superconductivity are mostly unknown.
In particular, the $\alpha$-Mn-type noncentrosymmetric Re$T$ ($T$ = Ti, Nb, Zr, and Hf) superconductors have been widely studied 
and demonstrated to show a superconducting state with broken TRS~\cite{Singh2014,Singh2017,Shang2018a,Shang2018b}. Yet, TRS seems to be 
preserved in the isostructural (but Re-free) Mg$_{10}$Ir$_{19}$B$_{16}$ 
and Nb$_{0.5}$Os$_{0.5}$~\cite{Acze2010,SinghNbOs}. 
Further, depending on synthesis protocol, Re$_3$W is either a 
centro- (hcp-Mg-type) or a noncentrosymmetric ($\alpha$-Mn-type) 
superconductor, yet neither is found to break 
TRS~\cite{Biswas2012}. 
In case of binary Re-Mo alloys, depending on the 
Re/Mo ratio, the compounds can exhibit up to four different crystal structures, including both cen\-tro\-sym\-me\-tric and non\-cen\-tro\-sym\-me\-tric cases. 
Most importantly, all these alloys become superconductors at low temperatures~\cite{Shang2019}. A comparative $\mu$SR study of Re-Mo alloys, 
covering all the different crystal structures, reveals that the 
spontaneous magnetic fields occurring below $T_c$ (an indication of TRS breaking) were 
only observed in elementary rhenium and in Re$_{0.88}$Mo$_{0.12}$. 
By contrast, TRS was preserved in the Re-Mo alloys with a lower 
Re-content (below $\sim 88$\%), independent of their centro- or noncentrosymmetric crystal structures~\cite{Shang2020ReMo}.  
Since both pure rhenium and Re$_{0.88}$Mo$_{0.12}$ have a simple (hcp-Mg-type) centrosymmetric structure, this strongly
suggests that a noncentrosymmetric structure and the accompanying ASOC effects are not essential in realizing the TRS breaking in Re$T$ superconductors. 
The $\mu$SR results regarding the Re-Mo family, as well as 
other Re-free $\alpha$-Mn-type superconductors, clearly imply 
that not only the Re presence, but also its amount are crucial for the appearance and the extent of TRS breaking in the Re$T$ superconductors. 
How these results can be understood within a more general framework requires further experimental and theoretical investigations.

This short review article focuses mostly on the experimental study of Re-based 
binary superconductors. In Sec.~\ref{sec:musr_intro}, we discuss the basic principles of our probe of choice, the $\mu$SR, here used 
to detect TRS breaking and to characterize the superconducting properties. 
Section~\ref{sec:Re_based_SC} describes the possible crystal 
structures and superconducting transition temperatures of Re$T$ binary 
alloys. In  Sec.~\ref{sec:Hc2_fields}, we focus on 
the upper critical fields and the 
order parameter in Re$T$ superconductors. Section~\ref{sec:TRS_breaking} discusses 
the TRS breaking in Re$T$ superconductors and its possible origins. 
Finally, in the last section, we outline some possible future research directions.

\begin{figure*}[!thp]
	\centering
	\includegraphics[width=0.8\textwidth,angle=0]{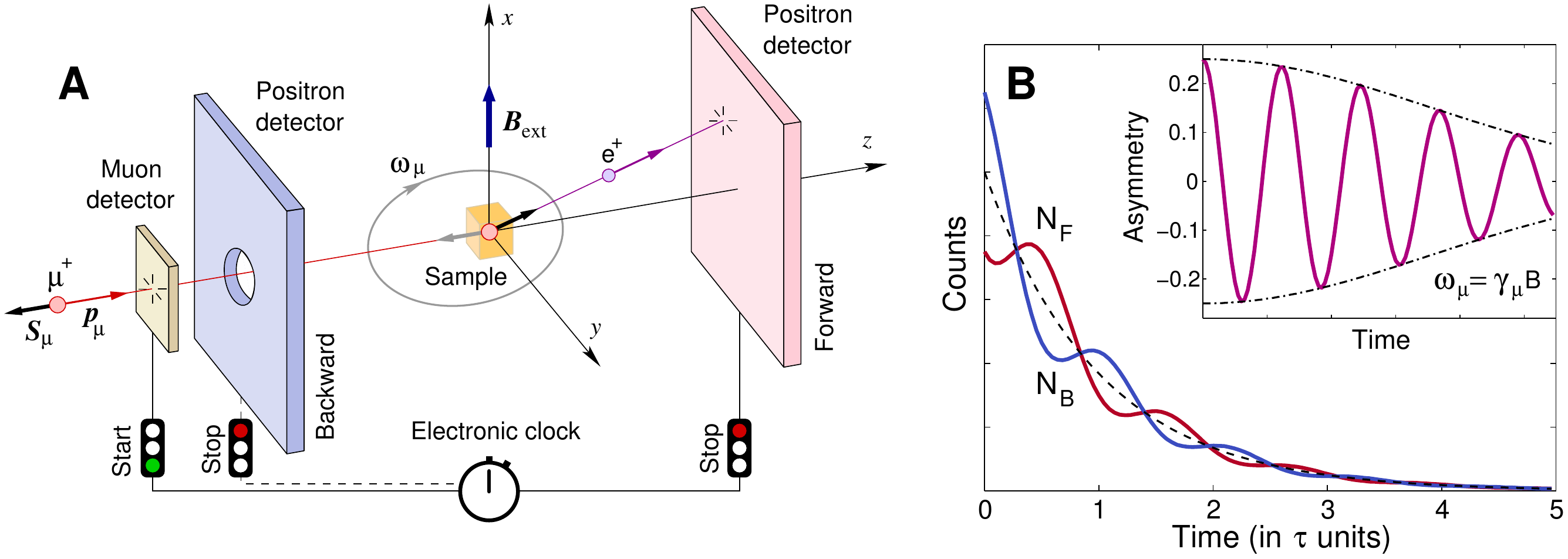}
	\caption{\label{fig:musr_basics}Principle of the time-differential $\mu$SR experiment. 
		\textbf{(A)} An incoming polarized muon (with spin $\boldsymbol{S_\mu}$ 
		antiparallel to momentum $\boldsymbol{p_\mu}$) is first detected by a 
		(thin) muon detector, which starts the electronic clock. 
		In the sample, the muon spin precesses in the internal/external 
		field until the muon decays into a positron.  
		This is emitted preferentially along the muon-spin direction 
		and hits one of the positron detectors [here forward (F) or backward (B)], 
		whose signal stops the clock. 
		The gray curve depicts the anisotropic positron-emission pattern at the 
		moment of muon implantation. The pattern rotates rigidly with the muon 
		spin (initially pointing towards the B detector) at an angular 
		frequency $\omega_{\mu} = \gamma_{\mu}B_\mathrm{ext}$. 
		\textbf{(B)} Detected positron counts in the F and B 
		detectors as a function of time after ca.\ $10^{7}$ events. 
		Inset: The asymmetry signal, obtained as the normalized difference 
		between the F and B counts, eliminates the inessential exponential 
		decay and highlights the signal decay (here assumed to be Gaussian) 
		reflecting the nature of the sample.
	}
\end{figure*}
%

\section{Muon-spin relaxation and rotation}
\label{sec:musr_intro}\enlargethispage{8pt}

Initially considered as an ``exotic'' technique, over the years muon-spin
rotation, relaxation, and resonance (known as $\mu$SR),
has become one of the most 
powerful methods to study the magnetic and superconducting properties
of matter. This follows from a series of fortunate circumstances, related
to the muon properties as a fundamental particle. Most notably, these 
include the 100\% initial muon-spin polarization, following the two-body 
decay from pions, and the subsequent preservation of such information 
through the weak decay into positrons. 
In the search for unconventional superconductivity, as well as for 
TRS breaking effects, the very high sensitivity of the $\mu$SR technique 
to tiny magnetic fields is especially important~\cite{Amato1997}.  
Below we briefly outline the basics of the $\mu$SR technique and direct 
the reader to other references for more detailed information
\citep{Blundell1999,Brewer2003,Yaouanc2011}.

\vspace{5mm}
\subsection{Principles of the $\mu$SR technique}
Central to the $\mu$SR method is the availability of polarized positive 
muon ($\mu^{+}$) beams, obtained by collecting the muons produced in 
the two-body decay of positive pions, 
$\pi^{+} \rightarrow \mu^{+} + \nu_{\mu}$ (with $\nu_{\mu}$ the muon 
neutrino), decaying at rest in the laboratory frame.
Since pions have no intrinsic angular momentum and neutrinos have 
a fixed helicity (relative orientation of spin and linear momentum), 
the resulting muon beam is 100\% spin polarized, with the muon spins 
directed antiparallel to the linear momentum (see \textbf{Figure~\ref{fig:musr_basics}A}).
Having an energy of $\sim$4.12\,MeV, muons can penetrate a sample between 
0.1\,mm and 1\,mm, depending on the sample density. 
Once implantated,
the monoenergetic muons decelerate 
within 100\,ps through ionization processes (which do not interact 
with the muon spin) and finally come to rest at an interstitial site, 
practically without loss of their initial spin polarization. 
From this moment, if subject to magnetic interactions, the muon-spin  
polarization $\boldsymbol{P}(t)$ evolves with time (the muon spin 
precesses around the local magnetic field), thus providing important 
information on the sample's magnetism. The detection of the $P(t)$ 
evolution is made possible by the parity-violating weak-decay interaction 
$\mu^{+} \rightarrow e^{+} + \nu_{e} + \bar\nu_{\mu}$ 
($e^{+}$, $\nu_{e}$, and $\bar\nu_{\mu}$ are the positron, electron neutrino, 
and muon antineutrino, respectively), which implies a preferential 
emission of positrons along the muon-spin 
direction at the time of decay (see \textbf{Figure~\ref{fig:musr_basics}A}, which 
depicts also the anisotropic positron-emission pattern). 
Thus, by detecting the spatial distribution of positrons as a 
function of time, one can determine the time evolution of the muon-spin 
polarization $\boldsymbol{P}(t)$.

A schematic diagram of a time-differential $\mu$SR experiment is shown 
in \textbf{Figure~\ref{fig:musr_basics}A}. The incoming muon triggers 
a clock that defines the starting time $t_0$. Once implanted, the muon 
spin precesses about the local magnetic field $\boldsymbol{B}(r)$ with 
a Larmor frequency $\omega_{\mu} = \gamma_{\mu}B(r)$, where 
$\gamma_{\mu}/2\pi = 135.53$\,MHz/T is the muon gyromagnetic ratio.
The clock stops when, after a mean lifetime of 2.197\,$\mu$s, the muon 
decays into a positron $e^+$, registered as an event by one of 
the positron detectors. The measured time intervals for ca.\ 10--50 
millions of such events are stored in a histogram, given by 
(see \textbf{Figure~\ref{fig:musr_basics}B}): 
\begin{equation}
	\label{eq:counts}
	N(t) = N_0\exp(-t/\tau_{\mu})[1 + A_0P(t)] + C. 
\end{equation}
Here, the exponential factor accounts for the radioactive muon decay, 
$N_0$ is the initial count rate at time $t_0$, while $C$ is a 
time-independent background (due to uncorrelated start and stop counts). 
As shown in the inset of \textbf{Figure~\ref{fig:musr_basics}B},
unlike the inessential exponential decay, the physical information in 
a $\mu$SR experiment is contained in the $A(t) = A_0P(t)$ term (often 
known as the $\mu$SR spectrum). 
Here, $A_0$ is the so-called initial asymmetry (typically 0.3, 
depending on the detector's solid angle and efficiency), 
while $P(t)$ is the muon-spin depolarization function, here given by 
the projection of $\boldsymbol{P}(t)$ on the unit vector describing 
the detector.
Since $P(t)$ represents the autocorrelation function of the 
muon spin $\boldsymbol{S}$, i.e.,  
$P(t) = \left< \boldsymbol{S}(t) \boldsymbol{S}(0)\right> /S(0)^2$, it 
depends on the average value, the distribution, and the time 
evolution of the internal magnetic fields, thus reflecting 
the physics of the 
magnetic interactions in the sample under study
To access the $\mu$SR signal we need to remove the extrinsic 
decay factor by combining the positron counts from pairs of 
opposite-lying detectors, for instance, $N_F$ and $N_B$ 
(for forward and backward), and obtain the asymmetry 
\tcr{
	$A(t) = [N_F(t) - \alpha N_B(t)]/ [N_F(t) + \alpha N_B(t)]$. Clearly, $A(t)$ 
	behaves as a normalized detector ``contrast'', proportional to $A_0$. The parameter $\alpha$ is introduced to take into account the different efficiencies  of the positron-detectors and needs to be determined by calibration.
}

\vspace{5mm}
\subsection{Transverse-field $\mu$SR}
\label{ssec:TF_muSR}
Depending on the reciprocal orientation of the external magnetic field 
$\boldsymbol{B}$ with respect to the initial muon-spin direction 
$\boldsymbol{S}(0)$, in a $\mu$SR experiment, two different configurations 
are possible. (i) In transverse-field (TF) $\mu$SR the externally 
applied field $\boldsymbol{B}$ is perpendicular to 
$\boldsymbol{S}(0)$ and the muon spin precesses 
around $\boldsymbol{B}$
(see \textbf{Figure~\ref{fig:musr_basics}A}). (ii) In a  
longitudinal field (LF) configuration the applied field is parallel to 
$\boldsymbol{S}(0)$, generally implying only a relaxing $\mu$SR signal. 

Although, in principle, the TF scheme shown in \textbf{Figure~\ref{fig:musr_basics}A} 
works fine, strong transverse fields perpendicular to the muon 
momentum ($\boldsymbol{p_{\mu}}$)
would deviate the muon beam too much from its original path. The Lorentz 
force can be zeroed by applying the field along the muon momentum. 
At the same time, to maintain the transverse geometry, the initial 
muon spin is rotated by $90^{\circ}$ (in the $x$ or $y$ direction) by 
using a so-called Wien filter, consisting of crossed electric and magnetic 
fields. Such a configuration is also known as transverse muon-spin mode,
while \textbf{Figure~\ref{fig:musr_basics}A} plots the longitudinal 
muon-spin mode (i.e., $\boldsymbol{p_{\mu}} \parallel \boldsymbol{S_{\mu}}$).

Since muons are uniformly implanted in the sample, they can detect 
the coexistence of different domains, characterized by distinct
$P_i(t)$ functions, whose amplitudes $A_i$ represent a measure of
the respective \emph{volume fractions}. In case of superconductors,
one can thus extract the SC volume fraction. More importantly, in a
TF-$\mu$SR experiment one can directly probe the SC flux-line lattice (FLL).
In this case, at the onset of superconductivity, the muon-spin
precession in a TF field loses coherence, reflecting the magnetic
field modulation (i.e., broadening) due to the FLL. 
The shape of the field distribution arising from
the FLL can be analyzed and eventually used to extract the magnetic
penetration depth $\lambda$ and the coherence length $\xi$
\cite{Sonier2000}. In many type-II superconductors, the simple relation,  
$\sigma_\mathrm{sc}^2 / \gamma_{\mu}^2 = 0.00371 \Phi_0^2/\lambda_\mathrm{eff}^4$,
connects the muon-spin depolarization rate in the superconducting 
phase, $\sigma_\mathrm{sc}$, with the effective magnetic penetration 
depth, $\lambda_\mathrm{eff}$ (here $\Phi_0$ is the magnetic flux 
quantum)~\cite{Barford1988,Brandt2003}.
In case of superconductors with relatively low upper critical 
fields, the effects of the overlapping vortex cores with increasing field ought to be
considered when extracting the magnetic penetration depth $\lambda_\mathrm{eff}$ from $\sigma_\mathrm{sc}$.
Since $\lambda(T)$ is sensitive to the low-energy excitations, its
evolution with temperature is intimately related to the structure
of the superconducting gap. Hence, $\mu$SR allows us to directly
study the \emph{symmetry} and \emph{value} of the superconducting gap.

More in detail, in a TF-$\mu$SR experiment, the time evolution of 
the asymmetry can be modeled by: 
\begin{equation}
	\label{eq:TF_muSR}
	A_\mathrm{TF}(t) = \sum\limits_{i=1}^n A_i \cos(\gamma_{\mu} B_i t + \phi) e^{- \sigma_i^2 t^2/2} +	A_\mathrm{bg} \cos(\gamma_{\mu} B_\mathrm{bg} t + \phi).
\end{equation}
Here $A_{i}$, $A_\mathrm{bg}$ and $B_{i}$, $B_\mathrm{bg}$ 
are the asymmetries and local fields sensed by implanted muons in the 
sample and the sample holder, $\gamma_{\mu}$
is the muon gyromagnetic ratio, $\phi$ is a shared initial phase, and $\sigma_{i}$ 
is the Gaussian relaxation rate of the $i$th component. The number of
required components is material dependent, typically in the $1 \leq n \leq 5$ range. In general, for superconductors with a large Ginzburg-Landau parameter
$\kappa$ ($\gg$ 1), the magnetic penetration depth is much larger than the coherence length. Hence, the field profiles of
each fluxon overlap strongly, implying a narrow field distribution. Consequently, a single-oscillating component
is sufficient to describe $A(t)$. In case of a small $\kappa$ ($\geq 1/\sqrt{2}$), 
the magnetic penetration depth is comparable with the coherence length.
The rather small penetration depth implies fast-decaying fluxon 
field profiles and a broad field distribution, in turn requiring 
multiple oscillations to describe $A(t)$. The choice of $n$ can be 
determined from the fast-Fourier-transform (FFT) spectra of the 
TF-$\mu$SR, which is normally used to evaluate the goodness of the fits.    
In case of multi-component oscillations, 
the first term in Equation~\eqref{eq:TF_muSR} describes the field distribution 
as the sum of $n$ Gaussian relaxations~\cite{Maisuradze2009}:
\begin{equation}
	\label{eq:TF_muSR_2}
	p(B) = \gamma_{\mu} \sum\limits_{i=1}^n \frac{A_i}{\sigma_i} \, \exp\!\left[-\frac{\gamma_{\mu}^2(B-B_i)^2}{2\sigma_i^2}\right].
\end{equation}
The first- and second moments of the field distribution in the sample 
can be calculated by: 
\begin{equation}
	\label{eq:2nd_moment}
	\langle B   \rangle = \sum \limits_{i=1}^n \frac{A_i B_i}{A_\mathrm{tot}}  
	\quad \textrm{and} \quad  
	\langle B^2 \rangle = \frac{\sigma_\mathrm{eff}^2}{\gamma_\mu^2} = \sum\limits_{i=1}^{n}%
	\frac{A_i}{A_\mathrm{tot}} \left[\frac{\sigma_i^2}{\gamma_{\mu}^2} + (B_i - \langle B \rangle)^2 \right],
\end{equation}
where $A_\mathrm{tot} = \sum_{i=1}^n A_i$. The total Gaussian relaxation 
rate $\sigma_\mathrm{eff}$ in Equation~\eqref{eq:2nd_moment} 
includes contributions from both a temperature-independent relaxation,  
due to nuclear moments ($\sigma_\mathrm{n}$), and a temperature-dependent 
relaxation, related to the FLL in the superconducting state ($\sigma_\mathrm{sc}$). 
The $\sigma_\mathrm{sc}$ values are then extracted by subtracting 
the nuclear contribution following 
$\sigma_\mathrm{sc}$ = $\sqrt{\sigma_\mathrm{eff}^{2} - \sigma^{2}_\mathrm{n}}$.

To get further insights into the superconducting gap value and its 
symmetry, the temperature-dependent superfluid density $\rho_\mathrm{sc}(T)$ 
[proportional to $\lambda_\mathrm{eff}^{-2}(T)$] is often analyzed by 
using a general model:
\begin{equation}
	\label{eq:rhos}
	\rho_\mathrm{sc}(T) = \frac{\lambda_0^2}{\lambda_\mathrm{eff}^2(T)} =  1 + 2\, \Bigg{\langle} \int^{\infty}_{\Delta_{k}} \frac{E}{\sqrt{E^2-\Delta_{k}^2}} \frac{\partial f}{\partial E} \mathrm{d}E \Bigg{\rangle}_\mathrm{FS}.
\end{equation}
Here, $\lambda_0$ is the effective magnetic penetration depth 
in the 0-K limit, $f = (1+e^{E/k_\mathrm{B}T})^{-1}$ is the Fermi function and $\langle \rangle_\mathrm{FS}$ represents an average over the Fermi surface~\cite{Tinkham1996}. 
$\Delta_{k}(T) = \Delta(T) A_{k}$ is an angle-dependent gap 
function, where $\Delta$ is the maximum gap value and $A_{k}$ is 
the angular dependence of the gap, equal to 1, $\cos2\phi$, and $\sin\theta$ 
for an $s$-, $d$-, and $p$-wave model, respectively, with $\phi$ 
and $\theta$ being the azimuthal angles.
The temperature dependence of the gap is assumed to follow 
the relation 
$\Delta(T) = \Delta_0 \mathrm{tanh} \{1.82[1.018(T_\mathrm{c}/T-1)]^{0.51} \}$~\cite{Tinkham1996,Carrington2003}, where $\Delta_0$ is the 0-K gap value.

\vspace{5mm}
\subsection{Zero-field $\mu$SR}
\label{ssec:ZF_muSR}
A particular case of LF, is that of zero-field (ZF) $\mu$SR, characterized  
by the absence of an external magnetic field. In this configuration 
the frequency of the $\mu$SR signal is exclusively proportional to
the internal magnetic field, making it possible to determined the size
of the ordered moments and, hence, the magnetic order parameter. 
Unlike various techniques, which require an external field to polarize 
the probe, $\mu$SR is unique in its capability of studying materials 
unperturbed by externally applied fields and in accessing their 
spontaneous magnetic fields. Due to the large muon magnetic moment
($\mu_{\mu} = 8.89\,\mu_\mathrm{N}$), ZF-$\mu$SR can sense
even very small internal fields ($\sim 10^{-2}$\,mT), hence
probing local magnetic fields of either nuclear or electronic nature.
In addition, since the muon is an elementary spin-1/2 particle, 
it acts as a purely magnetic probe, i.e., free of quadrupole interactions.
All these features, make ZF-$\mu$SR an ideal technique for detecting
TRS breaking in the superconducting state. This 
corresponds to the appearance (at the onset of SC) of spontaneous
magnetic moments, whose magnitude can be very small, often lacking
a proper magnetic order. As we show further on, in case of TRS breaking, 
we expect the appearance of an additional enhancement of
relaxation below $T_c$, 
reflecting the occurrence of these weak spontaneous fields. 
During the ZF-$\mu$SR measurements, to exclude the possibility of stray 
magnetic fields (typically larger than the weak internal fields), the 
magnets are quenched before starting the measurements, and an active 
field-nulling facility is used to compensate for stray fields down to 1\,$\mu$T.

If the amplitudes of the local fields reflect 
a Gaussian distribution with zero average 
(a rather common cir\-cum\-stance), 
the $\mu$SR signal consists of overlapping oscillations with 
different frequencies. While at short times the spin dephasing is limited, 
at long times it becomes relevant and gives rise to a so-called 
Kubo-Toyabe (KT) relaxation function~\cite{Yaouanc2011,Kubo1967}.
Two different models are frequently used to analyze the 
ZF-$\mu$SR data:
\begin{equation}
	\label{eq:KT_and_electr}
	A_\mathrm{ZF} = A_\mathrm{s}\left[\frac{1}{3} + \frac{2}{3}(1 -
	\sigma_\mathrm{ZF}^{2}t^{2} - \Lambda_\mathrm{ZF} t)\,
	\mathrm{e}^{\left(-\frac{\sigma_\mathrm{ZF}^{2}t^{2}}{2} - \Lambda_\mathrm{ZF} t\right)} \right] + A_\mathrm{bg},
\end{equation}
or 
\begin{equation}
	\label{eq:KT_and_electr_2}
	A_\mathrm{ZF} = A_\mathrm{s}\left[\frac{1}{3} + \frac{2}{3}(1 -
	\sigma_\mathrm{ZF}^{2}t^{2})\,
	\mathrm{e}^{-\frac{\sigma_\mathrm{ZF}^{2}t^{2}}{2}} \right] \mathrm{e}^{-\Lambda_\mathrm{ZF} t}  + A_\mathrm{bg}.
\end{equation}
Equation~\eqref{eq:KT_and_electr} is also known as a combined Gaussian- 
and Lorentzian Kubo-Toyabe function, with the additional exponential 
relaxation describing the electronic contributions present in 
many real materials. 
In polycrystalline samples, the 1/3-non-relaxing and the  
2/3-relaxing components of the asymmetry cor\-re\-spond to the powder 
average of the internal fields with respect to the initial 
muon-spin direction (statistically, with a 1/3 probability, the directions of the 
muon spin and of the local field coincide). Clearly, in case of single crystals, such prefactors might be different. 
The $\sigma_\mathrm{ZF}$ and $\Lambda_\mathrm{ZF}$ 
represent the zero-field Gaussian and Lorentzian relaxation rates, 
respectively. Typically, $\Lambda_\mathrm{ZF}$ shows an almost 
temperature-independent behavior. Hence, an increase of $\sigma_\mathrm{ZF}$ 
across $T_{c}$ can be attributed to the spontaneous magnetic fields 
which break the TRS, as e.g., in Re$T$~\cite{Singh2014,Singh2017,Shang2018b}. 
In case of diluted nuclear moments, $\sigma_\mathrm{ZF}$ is 
practically zero, hence, the TRS breaking is reflected in an 
increase of $\Lambda_\mathrm{ZF}$ below $T_c$, as e.g., 
in Zr$_3$Ir and CaPtAs~\cite{Shang2020,Shang2020b}.

\section{R\lowercase{e}-based superconductors}
\label{sec:Re_based_SC}
\begin{figure*}[tb]
	\centering
	\includegraphics[width=0.9\textwidth]{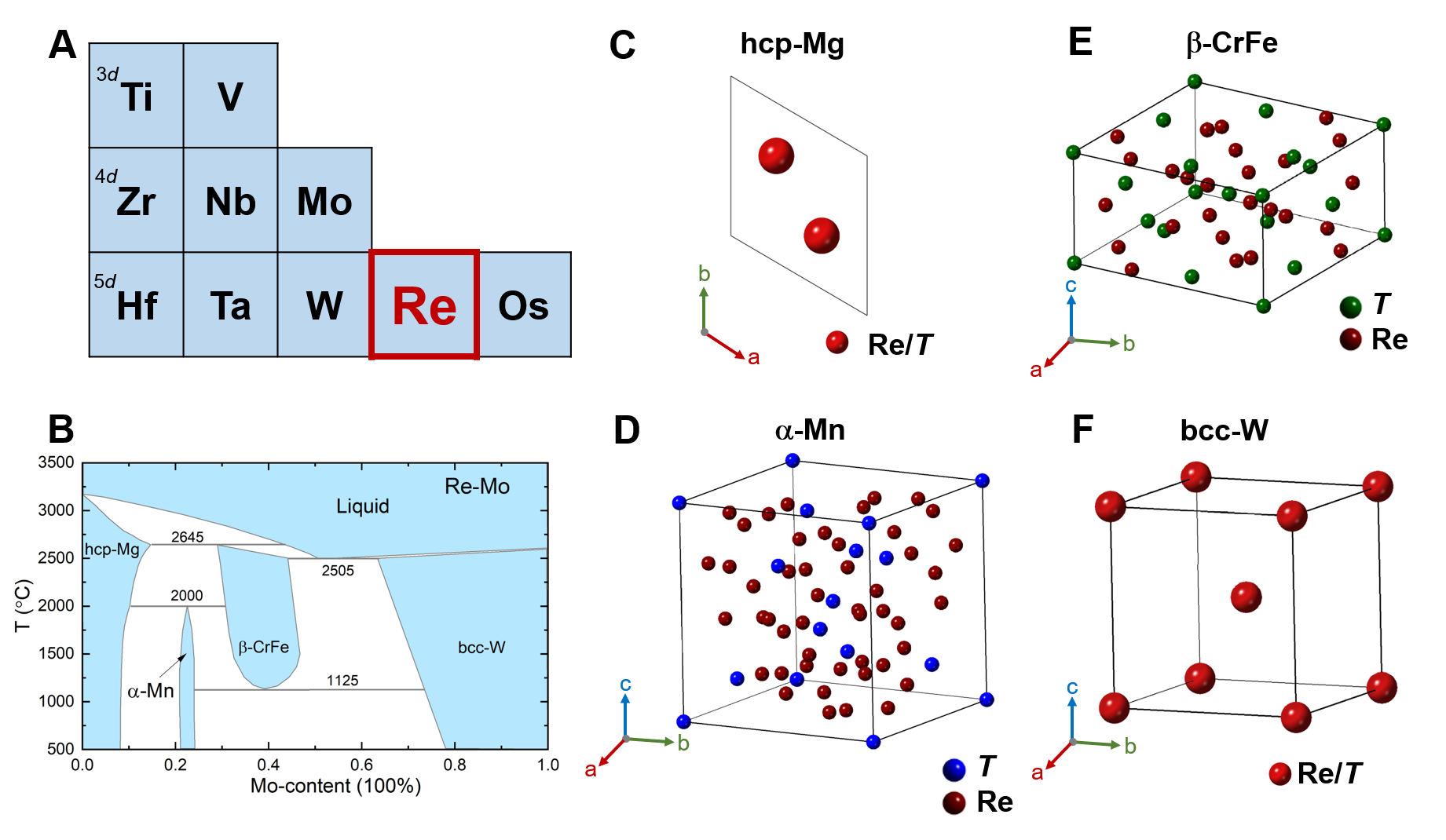}
	\caption{\label{fig:structure} Crystal structures of rhenium 
		transition-metal (Re$T$) superconductors. 
		\textbf{(A)} List of 3$d$, 4$d$, and 5$d$ early transition metals, 
		which can react with rhenium to form different crystal structures.
		\textbf{(B)} Binary phase diagram for the typical case 
		of Re-Mo alloys (data adopted from Ref.~\cite{Okamoto1996}). 
		\textbf{(C--F)} Unit cells of four most representative 
		crystal structures of Re$T$ binary compounds. Among these the 
		cubic $\alpha$-Mn type ($I\overline{4}3m$, No.\ 217) in \textbf{(D)} 
		is noncentrosymmetric, while the hexagonal hcp-Mg ($P6_3/mmc$, No.\ 194), tetragonal $\beta$-CrFe ($P4_2/mnm$, No.\ 136), and cubic bcc-W ($Im\overline{3}m$, No.\ 229) are centrosymmetric. 
		The atomic coordinates for each structure can be found in Refs.~\cite{Shang2019,Cenzual1986}.
	}
\end{figure*}

In this section, we review the different phases of the binary Re$T$ alloys.
These are obtained when rhenium reacts with various 
early transition metals (see \textbf{Figure~\ref{fig:structure}A}) and 
show rich crystal structures. Representative examples are shown in 
\textbf{Figures~\ref{fig:structure}C--F}, including the hexagonal hcp-Mg- ($P6_3/mmc$, No.\ 194), cubic $\alpha$-Mn- ($I\overline{4}3m$, No.\ 217), tetragonal $\beta$-CrFe- ($P4_2/mnm$, No.\ 136), and cubic bcc-W-type ($Im\overline{3}m$, No.\ 229).
Among these the cubic $\alpha$-Mn-type structure is noncentrosymmetric, while the rest are centrosymmetric~\cite{Okamoto1996}. 
Besides the above cases, a few other crystal structures have also been reported, including the  
cubic CsCl- ($Pm$-$3m$, No.\ 221), cubic Cr$_3$Si- ($Pm$-$3n$, No.\ 223), and trigonal Mn$_{21}$Zn$_{25}$-type ($R$-$3c$, No.\ 167)~\cite{Okamoto1996}. 
As for the pure elements listed in \textbf{Figure~\ref{fig:structure}A}, both Re and Os 
have an hcp-Mg-type structure, and show superconductivity below 2.7 and 0.7 K, respectively~\cite{Shang2018b,Roberts1976}; while V, Nb, Mo, Ta, and W all adopt a bcc-W-type structure, and become superconductors at
$\sim$ 5.4, 9.0, 1.0, 4.5, and 0.015 K, respectively~\cite{Roberts1976}. Unlike the above cases, Ti, Zr, and Hf can form either high-temperature bcc-W-type or low-temperature hcp-Mg-type structures, with $T_c$ $\sim$ 0.4, 0.6, and 0.13 K, respectively~\cite{Roberts1976}.

%

For $T$ = Ti 3$d$ metal, the known binary compounds are Re$_{24}$Ti$_5$, Re$_6$Ti, and ReTi~\cite{Murray1982,Philip1957}. 
The former two adopt a noncentroysmmetric $\alpha$-Mn-type structure and become 
superconductors below $T_c = 6$\,K~\cite{Shang2018a,Roberts1976,Singh2018}, while the latter one crystallizes in a cubic CsCl-type structure. To date, no detailed physical properties have been reported for ReTi. For $T$ = V, superconductivity has been reported in hcp-Mg-type Re$_{0.9}$V$_{0.1}$ ($T_c = 9.4$\,K), $\beta$-CrFe-type Re$_{0.76}$V$_{0.24}$ ($T_c = 4.5$\,K), and bcc-W-type Re$_{0.6}$V$_{0.4}$ ($T_c = 2.2$\,K)~\cite{Roberts1976,Jorda1986}. Also for them, to date a microscopic study of their SC is still missing. The cubic Cr$_3$Si-type Re$_{0.71}$V$_{0.29}$ has also been synthesized, but its physical properties were never characterized~\cite{Eremenko1990}. 

For $T$ = Zr 4$d$ metal, the $\alpha$-Mn-type Re$_{24}$Zr$_{5}$ ($T_c = 5$\,K) and Re$_6$Zr ($T_c = 6.7$\,K) have been investigated via both nuclear quadrupole resonance and $\mu$SR techniques~\cite{Singh2014,Matano2016}. Except for the $\alpha$-Mn-type Re-Zr alloys, the MgZn$_2$-type Re$_2$Zr (similar to hcp-Mg-type) and  Mn$_{21}$Zn$_{25}$-type Re$_{25}$Zr$_{21}$ have been synthesized~\cite{Okamoto1996}. Re$_2$Zr exhibits a $T_c$ value of $\sim 6$--7\,K~\cite{Roberts1976,Giorgi1970}, while Re$_{25}$Zr$_{21}$ has not been studied. For $T$ = Nb, depending on Re/Nb concentration, four different solid phases including hcp-Mg-, $\alpha$-Mn-, $\beta$-CrFe-, and bcc-W-type have been reported. On the Re-rich side, the hcp-Mg-type Re-Nb alloys are limited to less than 3\% Nb concentration~\cite{Okamoto1996}, whereas many $\alpha$-Mn-type Re-Nb binary alloys have been grown and widely studied by various techniques~\cite{Shang2018b,Cirillo2015,chen2013ReNb,Karki2011,Lue2011}, with the highest $T_c$ reaching 8.8\,K in Re$_{24}$Nb$_5$ (denoted as Re$_{0.82}$Nb$_{0.18}$ in the original paper~\cite{Shang2018b}). At intermediate Re/Nb values, for example, in $\beta$-CrFe-type Re$_{0.55}$Nb$_{0.45}$, $T_c$s in the range of 2 to 4\,K~\cite{Roberts1976} have been reported, but no microscopic studies yet. As for the Nb-rich side (Nb concentration larger than 60\%), here the Re-Nb alloys exhibit the same structure as that of pure Nb, but much lower 
$T_c$ values than Nb~\cite{Okamoto1996,Roberts1976}. For $T$ = Mo, the binary 
Re-Mo phase diagram (see \textbf{Figure~\ref{fig:structure}B}) covers also four different solid phases~\cite{Okamoto1996}. The binary Re-Mo alloys have been characterized by different techniques and all of them shown to become superconductors at low temperature~\cite{Shang2019,Shang2020ReMo}. 
The $T_c$ varies nonmonotonically upon changing the Mo concentration, 
giving rise to three distinct superconducting regions. On the Re-rich side, 
the first SC region shows the highest $T_\mathrm{c} \sim 9.4$\,K in 
the hcp-Mg-type Re$_{0.77}$Mo$_{0.23}$. The same material but with an 
$\alpha$-Mn-type structure can also be grown, with a $T_\mathrm{c}$ value about 1\,K lower than the hcp-Mg-type. In the second superconducting region, where the alloys adopt a $\beta$-CrFe-type structure, the superconducting transition temperature $T_\mathrm{c} \sim 6.3$\,K is almost independent of Mo content. Finally, on the Mo-rich side, all Re-Mo alloys display a cubic bcc-W-type structure and form a third superconducting region with the highest $T_\mathrm{c}$ reaching 12.4\,K in Re$_{0.4}$Mo$_{0.6}$. 

For $T$ = Hf 5$d$ metal, the Re-Hf alloys show a similar phase diagram to Re-Zr. 
With only $\sim$3\% Hf substitution, $T_c$ increases from $< 3$\,K to 7.3\,K in the hcp-Mg-type Re-Hf alloys~\cite{Roberts1976}. Both the $\alpha$-Mn-type Re$_6$Hf and the MgZn$_2$-type Re$_2$Hf become superconductors below $T_c \sim 6$\,K~\cite{Singh2017,Roberts1976,Giorgi1970,chen2016ReHf,Singh2016ReHf}, whereas the 
physical properties of Mn$_{21}$Zn$_{25}$-type Re$_{25}$Hf$_{21}$ 
remain largely unknown. On the Hf-rich side, the bcc-W-type alloys exhibit 
relatively low $T_c$s, e.g., $T_c = 1.7$\,K for Hf$_{0.875}$Re$_{0.125}$~\cite{Roberts1976}. 
For $T$ = Ta, although the
four different structures shown in \textbf{Figure~\ref{fig:structure}C--F} can be synthesized,
only the $\alpha$-Mn-type Re-Ta alloys have been well studied. For example, Re$_3$Ta and Re$_{5.5}$Ta show $T_c$ values of 4.7\,K and 8\,K, respectively~\cite{Barker2018,Arushi2020}.
On the Ta-rich side, the bcc-W-type Re-Ta alloys become superconducting at $T_c < 3.5$\,K, lower than the $T_c$ of pure Ta~\cite{Mamiya1970}. We note that in case of the $\beta$-CrFe-type Re-Ta alloys, no superconducting transition has been observed down to 1.8\,K in either Re$_{0.5}$Ta$_{0.5}$ or Re$_{0.6}$Ta$_{0.4}$. 
For $T$ = W, the Re-W alloys show a very similar phase diagram to Re-Mo in \textbf{Figure~\ref{fig:structure}B}. As the W concentration increases, the highest $T_c$ values 
reach $\sim 8$, 9, 6, and 5\,K in the hcp-Mg-, $\alpha$-Mn-, $\beta$-CrFe-, and the bcc-W-type alloys, respectively~\cite{Okamoto1996,Roberts1976}. 
Among them, only the hcp-Mg- and the $\alpha$-Mn-type Re$_3$W have 
been investigated~\cite{Biswas2012,Biswas2011}. 
Finally, in case of $T$ = Os, the Re-Os alloys show a rather 
monotonous phase diagram, since only hcp-Mg-type compounds
with $T_c$ values below 2\,K can be synthesized~\cite{Okamoto1996,Roberts1976}.

\section{Upper critical field and nodeless superconductivity}
\label{sec:Hc2_fields}
\begin{figure*}[tb]
	\centering
	\includegraphics[width=0.9\textwidth]{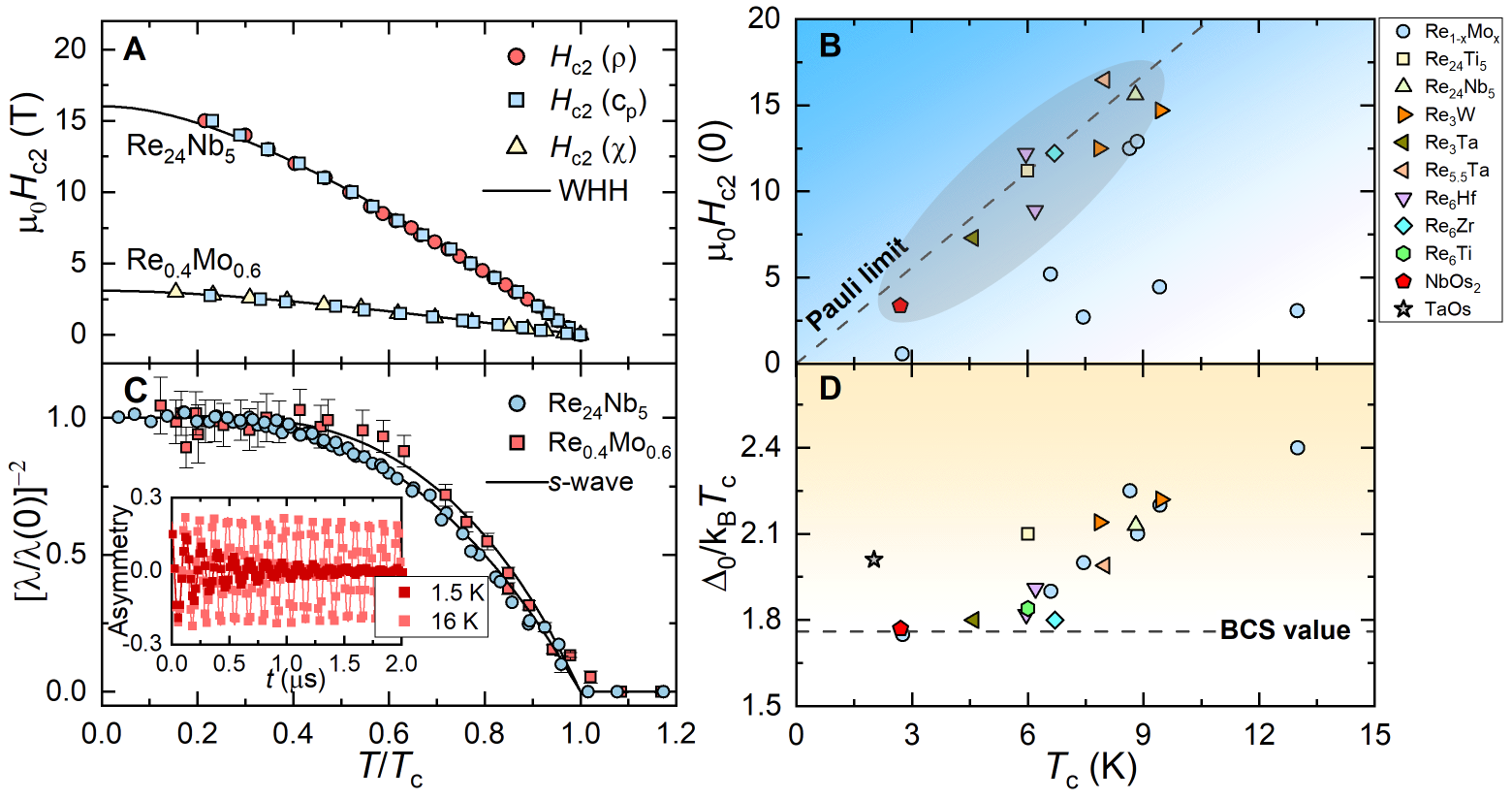}
	\caption{\label{fig:Hc2_gap} Upper critical field and superconducting energy gap. \textbf{(A)} The upper critical field $H_\mathrm{c2}$, as determined from electrical resistivity-, heat capacity-, and magnetic susceptibility measurements, as a function of the reduced superconducting transition temperature $T_c$/$T_c$(0) for Re$_{24}$Nb$_{5}$ and Re$_{0.4}$Mo$_{0.6}$. Solid-lines represent fits to the Werthamer-Helfand-Hohenberg (WHH) model. \textbf{(B)} Zero-temperature $H_\mathrm{c2}$ versus the superconducting transition temperature $T_c$ for $\alpha$-Mn-type NbOs$_2$ and all Re$T$ superconductors. 
		The shaded region in \textbf{(B)} marks the noncentrosymmetric $\alpha$-Mn type superconductors, while the dashed line indicates the Pauli  
		limit (i.e., $\mu_0 H_\mathrm{P} = 1.86 T_c$). 
		\textbf{(C)} Superfluid density vs reduced temperature $T_c$/$T_c$(0) for Re$_{24}$Nb$_{5}$ and Re$_{0.4}$Mo$_{0.6}$. Lines are fits to a fully-gapped $s$-wave model. 
		The insert shows the TF-$\mu$SR spectra for Re$_{0.4}$Mo$_{0.6}$ measured 
		in a field of 60\,mT in the normal- (16\,K) and the superconducting state (1.5\,K). 
		Solid lines are fits to Equation~\ref{eq:TF_muSR}.  	
		\textbf{(D)} Zero-temperature superconducting energy gap $\Delta_0$ (in k$_\mathrm{B}T_c$ units) as a function of $T_c$ for Re$T$ and $\alpha$-Mn-type NbOs$_2$ and TaOs superconductors. Here, the dashed line     
		represents the BCS superconducting gap in the weak-coupling 
		limit (i.e., 1.76\,$k_\mathrm{B}T_c$). Data were taken from Refs.~\cite{Singh2014,Singh2017,Shang2018a,Shang2018b,Biswas2012,Shang2020ReMo,Singh2018,chen2016ReHf,Singh2016ReHf,Arushi2020,Biswas2011,Singh2019NbOs2,SinghTaOs,mayoh2017ReZr}.
			}
\end{figure*}

As mentioned in the introduction, due to the mixture of singlet- 
and triplet paring, some NCSCs may ex\-hib\-it relatively high 
upper critical fields, often very close to or even exceeding
the Pauli limit, as e.g., 
CePt$_3$Si~\cite{Bauer2004}, Ce(Rh,Ir)Si$_3$~\cite{Kimura2007,Sugitani2006}, 
and recently (Ta,Nb)Rh$_2$B$_2$~\cite{Carnicom2018}. 
Therefore, the upper critical field can provide valuable clues about 
the nature of superconductivity.
To investigate the temperature evolution of the upper critical 
field $H_{c2}(T)$, in general, the temperature- (or field-) dependent 
electrical resistivity $\rho$, magnetic susceptibility $\chi$, and specific heat $C/T$ at various magnetic fields (or at various temperatures) are measured~\cite{Shang2018a,Shang2018b,Shang2020ReMo}. 
As an example, \textbf{Figure~\ref{fig:Hc2_gap}A} shows the $H_{c2}(T)$ for Re$_{24}$Nb$_5$ ($\alpha$-Mn-type) and Re$_{0.4}$Mo$_{0.6}$ (bbc-W-type) versus 
the normalized temperature $T/T_c(0)$. To obtain the upper 
critical field in the zero-temperature limit, $H_{c2}(0)$, the Werthamer-Helfand-Hohenberg (WHH) or the Ginzburg-Landau (GL) models are usually applied when analyzing the 
$H_{c2}(T)$ data of Re$T$ superconductors. Both models can adequately describe single-gap superconductors. Here, in case of Re$_{24}$Nb$_5$ and Re$_{0.4}$Mo$_{0.6}$, the WHH model (solid line in \textbf{Figure~\ref{fig:Hc2_gap}A}) reproduces the data very well and gives $\mu_0H_{c2}(0)$ = 15.6\,T, and 3.08\,T, respectively. \textbf{Figure~\ref{fig:Hc2_gap}B} summarizes the $\mu_0H_{c2}(0)$ values of the Re$T$ and $\alpha$-Mn-type NbOs$_2$ superconductors. As discussed in Section~\ref{sec:Re_based_SC}, most of the previous studies focused exclusively on $\alpha$-Mn-type Re$T$ 
superconductors, the physical properties of the other Re$T$ 
superconductors being practically neglected  and requiring 
further studies. Unlike other Re$T$, all Re-Mo alloys belonging to four 
different structures have been studied via macro- and 
microscopic techniques~\cite{Shang2019,Shang2020ReMo}. The $\mu_0H_{c2}(0)$ of centrosymmetric Re-Mo alloys, including hcp-Mg-, $\beta$-CrFe-, and bcc-W-type, are far away from the Pauli limit $\mu_0 H_\mathrm{P} = 1.86\,T_{c}$ (indicated by a dashed line in \textbf{Figure~\ref{fig:Hc2_gap}B}). Conversely, the $\alpha$-Mn-type Re$T$ and 
NbOs$_2$ both exhibit large upper critical fields, very close to or even exceeding the Pauli limit, despite their different $T_c$ values. For example, $\mu_0 H_{c2}(0) =  15.6$ and 16.5\,T for Re$_{24}$Nb$_5$ and Re$_{5.5}$Ta, 
while their $\mu_0 H_P(0)$ are 16.4 and 14.9\,T, respectively.
The hcp-Mg-type Re$_3$W also exhibits a relatively high $H_{c2}$, as 
determined from electrical resistivity data. However, its
$H_{c2}$ value might be 
overestimated since, e.g., at 9\,T, no zero resistivity could be observed down 
to 2\,K. Therefore, other bulk techniques, including magnetization- or heat capacity measurements are required to determine the intrinsic $H_{c2}$. 
In general, it would be interesting to know the $H_{c2}$ values of other centrosymmetric Re$T$ superconductors. Overall, the upper critical fields in \textbf{Figure~\ref{fig:Hc2_gap}B} indicate the possibility of singlet-triplet mixing 
in the noncentrosymmetric $\alpha$-Mn-type superconductors.

%

Transverse-field $\mu$SR represents one of the most powerful techniques to investigate the superconductivity at a microscopic level. 
To illustrate this, in the inset of \textbf{Figure~\ref{fig:Hc2_gap}C} 
we show two typical TF-$\mu$SR spectra for bcc-W-type Re$_{0.4}$Mo$_{0.6}$ in the normal and the superconducting states. Below $T_c$, the fast decay induced by FLL (encoded into $\sigma_\mathrm{sc}$) is clearly visible, while the slow decay in the normal state is attributed to the randomly oriented nuclear magnetic moments. By comparing the two spectra, one can also determine 
the superconducting volume fraction of a superconductor. 
As an example, the main panel of \textbf{Figure~\ref{fig:Hc2_gap}C} shows the normalized superfluid density calculated from $\sigma_\mathrm{sc}(T)$, which is proportional to $[\lambda(T)/\lambda(0)]^{-2}$ (see details in Section~\ref{ssec:TF_muSR}),  
as a function of the reduced temperature $T/T_\mathrm{c}(0)$ for Re$_{24}$Nb$_5$ and Re$_{0.4}$Mo$_{0.6}$~\cite{Shang2018b,Shang2020ReMo}. The low-$T$ superfluid density is practically independent of temperature, clearly suggesting a lack of low-energy excitations and a fully-gapped superconductivity. Contrarily, such excitations exist in case 
of nodes in the superconducting gap, implying a temperature-dependent 
superfluid density 
below $\sim T_c/3$. As shown by solid lines in \textbf{Figure~\ref{fig:Hc2_gap}C}, the $\rho_\mathrm{sc}(T)$ of Re$T$ superconductors is
described very well by a fully-gapped $s$-wave model (see Equation~\ref{eq:rhos}). The other $\alpha$-Mn-type Re$T$, TaOs, and NbOs$_2$ exhibit similar temperature-invariant superfluid densities~\cite{Singh2014,Singh2017,Shang2018a,Biswas2012,Singh2018,Arushi2020,Singh2019NbOs2,SinghTaOs}.
Although Re$T$ alloys adopt different crystal structures (i.e., centrosymmetric or noncentrosymmetric, see \textbf{Figure~\ref{fig:structure}C--F}) and have different $T_c$ values, they regularly exhibit low-$T$ superfluid densities which are independent of temperature~\cite{Shang2020ReMo}. 
Except for $T$ = Mo (and for some $\alpha$-Mn structures), a systematic 
microscopic study of superconductivity in other Re$T$ superconductors 
is still missing. Clearly, it would be interesting to know if 
their SC behavior is similar to that of Re-Mo alloys. 
The nodeless SC scenario in Re$T$ alloys is also supported by other 
techniques, as the electronic specific heat, the magnetic penetration 
depth measured via the tunnel-diode-oscillator-based technique, 
or the point-contact Andreev reflection~\cite{Shang2018a,Shang2018b,Shang2020ReMo,chen2016ReHf,Singh2016ReHf,Biswas2011,mayoh2017ReZr,Pang2018,Parab2019}. In addition, some studies have found evidence of 
two-gap superconductivity in Re$_{0.82}$Nb$_{0.18}$ and Re$_6$Zr~\cite{Cirillo2015,Parab2019}.

\textbf{Figure~\ref{fig:Hc2_gap}D} summarizes the zero-temperature superconducting energy gap value for Re$T$ and $\alpha$-Mn-type NbOs$_2$ and TaOs superconductors as a function of their critical temperatures. Most of them 
exhibit a $\Delta_0$/k$_\mathrm{B}T_c$ ratio larger than 1.76 (see dashed line in \textbf{Figure~\ref{fig:Hc2_gap}D}), the value expected for a weakly coupled BCS superconductor, which indicates a moderately strong coupling in these superconductors. In addition, the specific-heat discontinuity at $T_c$ (i.e., $\Delta C/\gamma T_c$) is larger than the conventional BCS value of 1.43, again indicating an enhanced electron-phonon coupling~\cite{Shang2018a,Shang2018b,Shang2020ReMo,chen2016ReHf,Singh2016ReHf,Biswas2011,mayoh2017ReZr}. 
As mentioned above, it is worth noting that the superconducting parameters of all the other Re$T$ materials (except for $\alpha$-Mn-type 
and $T$ = Mo) are missing, prompting further research efforts in this direction. 

As discussed in the introduction, the lack of inversion symmetry in the 
NCSCs often induces an ASOC. This splits the Fermi surface by 
lifting the degeneracy of the conduct{ion electrons, thus 
	allowing admixtures of spin-singlet and spin-triplet pairing. 
	In general, the strength of ASOC determines the degree of such an admixture and thus the superconducting properties of NCSCs~\cite{Bauer2012,Smidman2017}.
	A fully-gapped superconductor (i.e., dominated by spin-singlet
	pairing) can be tuned into a nodal superconductor (dominated by
	spin-triplet pairing) by increasing the strength of ASOC. Such mechanism has been successfully demonstrated, e.g., in weakly-correlated Li$_2$Pt$_3$B ($E_\mathrm{SOC}$/k$_\mathrm{B}T_c$ $\sim 831$)~\cite{yuan2006,NiShiyama2007}, CaPtAs ($E_\mathrm{SOC}$/k$_\mathrm{B}T_c$ $\sim 800$)~\cite{Shang2020,Xie2019}, and in strongly-correlated CePt$_3$Si ($E_\mathrm{SOC}$/k$_\mathrm{B}T_c$ $\sim 3095$) superconductors~\cite{Bauer2004,Samokhin2004}, all exhibiting a relatively large band splitting $E_\mathrm{SOC}$ compared to their superconducting energy scale k$_\mathrm{B}T_c$. 
	In the $\alpha$-Mn-type Re$T$ alloys, the density of states (DOS) near the Fermi level  
	is dominated by the 5$d$ orbitals of rhenium atoms, while contributions 
	from the $d$ orbitals of $T$ atoms are negligible~\cite{Suetin2013,Winiarski2014,Mojammel2016}.
	Therefore, a possible enhancement of SOC due to 3$d$-(e.g., Ti, V) 
	up to 5$d$-electrons (e.g., Hf, Ta, W, Os) will, in principle, neither increase the band splitting $E_\mathrm{SOC}$ nor
	affect the pairing admixture and thus the superconducting properties 
	of $\alpha$-Mn-type Re$T$ superconductors. According to band-structure calculations,  
	in Re$_6$Zr, the SOC-induced band splitting 
	is about 30\,meV~\cite{Mojammel2016}, implying a very small ratio $E_\mathrm{SOC}$/k$_\mathrm{B}T_c$ $\sim 25$, 
	comparable to that of fully-gapped Li$_2$Pd$_3$B, Mo$_3$P, and Zr$_3$Ir superconductors~\cite{Shang2020b,NiShiyama2007,Shang2019b}. Therefore, despite the relatively 
	large SOC of rhenium atoms, its effects are too weak 
	to significantly influence the bands near the Fermi level. 
	This might explain why all the $\alpha$-Mn-type Re$T$ superconductors 
	exhibit nodeless superconductivity, more consistent with a spin-singlet 
	dominated pairing~\cite{Singh2014,Singh2017,Shang2018a,Shang2018b,Biswas2012,Barker2018}. 
	However, we recall that often, due to the similar magnitude and same-sign of the order parameter on the spin-split Fermi surfaces, a possible mixed-pairing superconductor
	may be challenging to detect
	or to distinguish from a single-gap $s$-wave superconductor~\cite{Yip2014}. The almost spherical symmetry of the Fermi surface in these materials may also explain their BCS-like superconducting states~\cite{Winiarski2014}. 
	As for the other centrosymmetric Re$T$ alloys, in most of them 
	the Re and $T$ atoms occupy the same atomic positions in the unit cell. 
	In this case, as the $T$-content increases, the contribution of $T$ $d$ orbitals to the DOS is progressively enhanced, at the expense of the Re $5d$ orbitals.  
	Therefore, the chemical substitution of Re by another 3$d$, 4$d$, 
	or 5$d$ $T$ metal (see \textbf{Figure~\ref{fig:structure}}), 
	should significantly tune the SOC and, hence, the band splitting, 
	an interesting hypothesis waiting for theoretical confirmation.
	However, even for $T$ = Hf, Ta, W, and Os, the maximum $E_\mathrm{SOC}$ 
	should still be comparable to that of $\alpha$-Mn-type Re$T$ alloys. 
	Finally, irrespective of the strength of SOC, due to their 
	centrosymmetric crystal structures, these compounds may exhibit 
	either singlet- or triplet-pairing, but not an admixture of both. 
	According to the TF-$\mu$SR results (see \textbf{Figure~\ref{fig:Hc2_gap}C}), 
	despite a change in SOC and of the different crystal structures 
	(see \textbf{Figure~\ref{fig:structure}C--F}), all Re$T$ 
	superconductors exhibit fully-gapped superconducting states. 
	This finding strongly suggests that, in the Re$T$ superconductors, 
	spin-singlet pairing is dominant.

\section{Time-reversal symmetry breaking}
\label{sec:TRS_breaking}
%
%
\begin{figure*}[tb]
	\centering
	\includegraphics[width=0.8\textwidth]{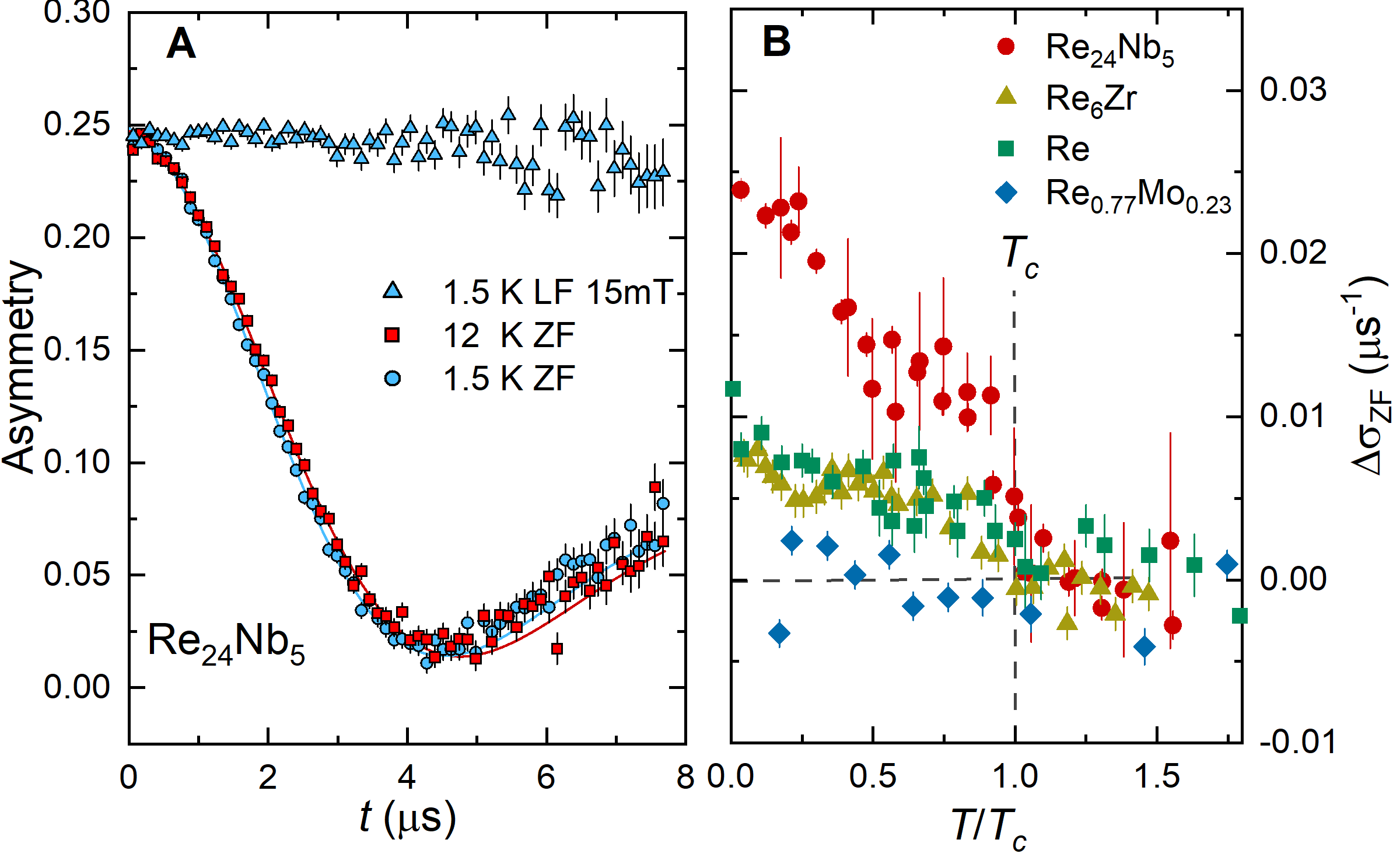}
	\caption{\label{fig:ZF_muSR} ZF-$\mu$SR and evidence for TRS breaking. 
		\textbf{(A)} ZF-$\mu$SR spectra for Re$_{24}$Nb$_{5}$ collected in the superconducting and normal states. Top: additional 
		$\mu$SR data collected at 1.5\,K in a 15-mT longitudinal field, are also shown. The solid lines are fits using Equation~\ref{eq:KT_and_electr}. 
		\textbf{(B)} Gaussian relaxation rate $\Delta\sigma_\mathrm{ZF}$ 
		vs $T$/$T_c$ for Re$_{24}$Nb$_5$, Re$_6$Zr, Re, and Re$_{0.77}$Mo$_{0.23}$  
		--- here $\Delta\sigma_\mathrm{ZF}(T) = \sigma_\mathrm{ZF}(T) - \sigma_\mathrm{ZF}(T > T_c)$. 
		While for the first three there is a clear increase of $\Delta\sigma_\mathrm{ZF}$ 
		across $T_c$ (hence a breaking of TRS), no changes occur in the last 
		case (TRS is preserved).  
		Data were taken from Refs.~\cite{Singh2014,Shang2018a,Shang2018b,Shang2020ReMo}.
	}
\end{figure*}

Owing to its very high sensitivity (see details in Section~\ref{ssec:ZF_muSR}), 
ZF-$\mu$SR has been successfully used to search for spontaneous magnetic fields, reflecting the breaking of TRS in the superconducting states of different types of superconductors, as 
e.g., Sr$_2$RuO$_4$, UPt$_3$, PrOs$_4$Sb$_{12}$, LaNiGa$_2$, LaNiC$_2$, La$_7$(Rh,Ir)$_3$, and $\alpha$-Mn-Re$T$~\cite{Hillier2009,Barker2015,Singh2014,Singh2017,Shang2018a,Shang2018b,Singh2018La7Rh3,Luke1993,Luke1998,Xia2006,Aoki2003,Schemm2014,Hillier2012}. 
The latter three are typical examples of weakly-correlated NSCSs, to be contrasted with strongly-correlated NCSCs, 
where either the TRS is broken by a coexisting long-range magnetic order, 
or the tiny TRS-breaking signal is very difficult to detect due to the presence of strong magnetic fluctuations~\cite{Amato1997}. In the former case, the broken TRS is unrelated to
the superconductivity, while in the later case, a genuine
TRS breaking  effect is masked by the much faster muon-spin relaxation caused by magnetic fluctuations. Therefore, in general, a TRS breaking effect is 
more easily (and reliably) detected in weakly-correlated- or non-magnetic 
superconductors using $\mu$SR techniques.  
Normally, in the absence of external fields, the onset of superconductivity does not imply changes in the ZF-$\mu$SR relaxation rate. However, in presence of a 
broken TRS, the onset of a tiny spontaneous polarization 
or of currents gives rise to associated (weak) magnetic fields, readily detected by ZF-$\mu$SR as an increase in the relaxation rate. Given the tiny size of such effects, the ZF-$\mu$SR measurements are usually performed in both the normal- and the superconducting state with a relatively high statistics, at least twice that of the TF-$\mu$SR spectra. 
As an example, \textbf{Figure~\ref{fig:ZF_muSR}A} plots the ZF-$\mu$SR spectra of $\alpha$-Mn-type 
Re$_{25}$Nb$_5$, 
with the other Re$T$ superconductors showing a similar behavior. The ZF-$\mu$SR spectra collected below- and above $T_c$ (at 1.5 and 12\,K) exhibit small yet measurable differences. 
The lack of any oscillations in the spectra, implies the non-magnetic nature of Re$T$ superconductors.
Further, longitudinal-field $\mu$SR measurements under a relatively small applied field (typically a few tens of mT) in the superconducting state are usually performed to check if the applied field can fully decouple the muon spins from the weak spontaneous magnetic fields, and thus exclude extrinsic effects.   
In non-magnetic materials in the
absence of external magnetic fields, the muon-spin relaxation is mostly determined by the interaction between the muon spins and the randomly oriented nuclear magnetic moments. 
Therefore, the spontaneous magnetic fields due to the 
TRS breaking will be reflected in an additional increase of muon-spin relaxation.  
The ZF-$\mu$SR asymmetry can be described by means of a Gaussian- or a 
Lorentzian Kubo-Toyabe relaxation, or a combination thereof (see Equations~\ref{eq:KT_and_electr} and \ref{eq:KT_and_electr_2}). 
\textbf{Figure~\ref{fig:ZF_muSR}B} summarizes the Gaussian relaxation 
rate $\sigma_\mathrm{ZF}$ versus the reduced temperature $T/T_c$ for the
$\alpha$-Mn-type Re$_{24}$Nb$_5$, Re$_6$Zr, and Re$_{0.77}$Mo$_{0.23}$, and the hcp-Mg-type elementary Re. Above $T_c$, all the samples show a temperature-independent 
$\sigma_\mathrm{ZF}$. Except for Re$_{0.77}$Mo$_{0.23}$, a small yet clear 
increase of $\sigma_\mathrm{ZF}(T)$ below $T_c$ indicates the onset of spontaneous magnetic fields, which represent the signature of TRS breaking in the superconducting state~\cite{Singh2014,Shang2018b,Shang2020ReMo}. 
The other $\alpha$-Mn-type superconductors, e.g., Re$_6$Ti, and Re$_6$Hf~\cite{Singh2017,Singh2018}, show similar $\sigma_\mathrm{ZF}(T)$ to Re$_{24}$Nb$_5$ and Re$_6$Zr, and thus the breaking of TRS in the superconducting state. At the same time, in the isostructural Re$_3$Ta, Re$_{5.5}$Ta, and Re$_3$W cases, there is no clear increase in $\sigma_\mathrm{ZF}(T)$ upon crossing $T_c$, implying a preserved TRS~\cite{Biswas2012,Barker2018,Arushi2020}.

Recently, the breaking of TRS and the presence of nodes in the SC gap, attributed to an admixture of singlet- and triplet paring, 
has been reported in the noncentrosymmetric CaPtAs superconductor~\cite{Shang2020}. 
In general, however, the breaking of TRS in the superconducting 
state and a lack of space-inversion symmetry in 
the crystal structure are independent events, 
not necessarily occurring together. For instance, the unconventional 
spin-triplet pairing is expected to break TRS below $T_c$, as 
has been shown, e.g., in Sr$_2$RuO$_4$, UPt$_3$, and UTe$_2$ triplet superconductors~\cite{Ran2019,Luke1993,Luke1998,Xia2006,Schemm2014,Ishida1998,Tou1998,Mackenzie2003,Joynt2002}.
An $s + id$ spin-singlet state was proposed to account for the TRS breaking in some iron-based high-$T_\mathrm{c}$ superconductors~\cite{Lee2009}, where a nodal gap 
is also expected. The frequent occurrence of TRS breaking in the fully-gapped 
(i.e., dominated by spin-singlet pairing) Re$T$ superconductors 
(see Section~\ref{sec:Hc2_fields}) is, therefore, rather puzzling.
A similarly surprising result was the report that elementary
rhenium also exhibits signatures of TRS breaking in its superconducting state 
(see \textbf{Figure~\ref{fig:ZF_muSR}B}), with $\Delta\sigma_\mathrm{ZF}(T)$ 
being comparable to that of Re$_6$Zr~\cite{Shang2018b,Shang2020ReMo}.  
Since elementary rhenium adopts a centrosymmetric hcp-Mg crystal 
structure (see \textbf{Figure~\ref{fig:structure}C}), this indicates 
that a lack of inversion symmetry and the accompanying ASOC effects are not 
crucial factors for the occurrence of TRS breaking in Re$T$ superconductors. 
Further on, a comparison of ZF-$\mu$SR measurements on Re-Mo alloys 
with different Re/Mo contents, covering almost all the crystal structures 
reported in \textbf{Figure~\ref{fig:structure}}, shows that only Re and Re$_{0.88}$Mo$_{0.12}$ exhibit a broken TRS in the superconducting state, while those with a 
higher Mo-content ($\sim$23\%--60\%), including both the centrosymmetric- and noncentrosymmetric Re$_{0.77}$Mo$_{0.23}$, preserve the TRS. Considering the preserved TRS in Mg$_{10}$Ir$_{19}$B$_{16}$, and Nb$_{0.5}$Os$_{0.5}$~\cite{Acze2010,SinghNbOs},
all of which share the same 
$\alpha$-Mn-type structure, this implies that TRS breaking in Re$T$ superconductors is clearly not related to the noncentrosymmetric crystal structure or to a
possible mixed pairing but, most likely, is due to the presence of rhenium and to its amount.  
Such conclusion is further reinforced by the preserved TRS in many Re-based 
superconductors, whose Re-content is below a certain threshold. Such cases include, e.g., Re$_3$W, Re$_3$Ta, Re-Mo (with Mo-content higher than 12\%)~\cite{Biswas2012,Shang2020ReMo,Barker2018}, the recently reported Re-B superconductors~\cite{Sharma2020}, and the diluted ReBe$_{22}$ superconductor~\cite{Shang2019c}. 
Moreover, by comparing the ZF-$\mu$SR relaxation across various 
Re$T$ superconductors, a clear positive correlation between 
$\Delta\sigma_\mathrm{ZF}$ (i.e., spontaneous fields) and the size of 
the nuclear magnetic moments $\mu_\mathrm{n}$ was identified~\cite{Shang2018b}. For instance, among the Re$T$ superconductors, Re$_{24}$Nb$_5$ 
shows the largest spontaneous fields below $T_c$ 
(see \textbf{Figure~\ref{fig:ZF_muSR}B}), a fact compatible with 
the large nuclear magnetic moment of 
niobium, practically twice that of rhenium (6.17 vs.\ 3.2\,$\mu_\mathrm{N}$). 
However, the correlation between $\mu_\mathrm{n}$ and $\Delta\sigma_\mathrm{ZF}$ 
alone cannot explain TRS breaking, considering that elementary Nb itself,  
despite having the highest $\mu_\mathrm{n}$, does not break TRS. 
Clearly, the origin of such correlation is not yet understood and 
it requires further experimental and theoretical studies. 

If SOC can be ignored, an alternative mechanism, which can 
account for the TRS breaking in Re$T$ superconductors in presence of 
a fully-opened superconducting gap, is the internally-antisymmetric 
nonunitary triplet (INT) pairing. 
The INT pairing was originally proposed to explain the TRS breaking and nodeless SC in centrosymmetric LaNiGa$_2$~\cite{Hillier2012, Weng2016,Ghosh2020} 
and noncentrosymmetric LaNiC$_2$~\cite{Hillier2009, Quintanilla2012}, 
both exhibiting a relatively weak SOC. 
In case of INT pairing, the superconducting pairing function is antisymmetric with respect to the orbital degree of freedom, while remaining symmetric in the spin- 
and crystal-momentum channels~\cite{Hillier2009,Hillier2012, Weng2016,Quintanilla2012}.
Since in Re$T$ superconductors, too, the SOC interaction is relatively  
weak ($\sim30$\,meV, see Section~\ref{sec:Hc2_fields})~\cite{Mojammel2016} and since neither TRS breaking nor 
the nodeless SC are related to the symmetry of Re$T$ crystal structures, 
the effect of SOC to the observed TRS breaking is insignificant. 
This could, therefore, explain why a lack of inversion symmetry (essential to SOC) is not a precondition for TRS breaking in Re$T$ superconductors. 
Moreover, the occurrence of an INT state relies on the availability of a local-pairing mechanism driven by Hund's rules, e.g., by Ni $3d$-electrons 
in LaNiC$_2$ and LaNiGa$_2$~\cite{Hillier2009,Hillier2012,Weng2016,Ghosh2020,Quintanilla2012}. Such local-pairing mechanism may also occur in Re$T$ superconductors, 
since rhenium too can be magnetic~\cite{Yang2015,Kochat2017}. 
This consideration is also in good agreement with the observation 
that TRS breaking depends on Re content, but not on a noncentrosymmetric 
crystal structure.   
		
\section{Conclusion}
In this short review we focused on recent experimental studies of
Re$T$ superconductors, where time-reversal symmetry breaking 
effects are often present and whose superconductivity 
can, therefore, be considered as unconventional. Due to its high 
sensitivity to the weak internal fields associated with TRS
breaking, $\mu$SR represents one of the key techniques in the search 
for TRS-breaking effects in the superconducting state. 
Nonetheless, in certain cases, the amplitude of the spontaneous magnetic fields 
(the fingerprint of TRS breaking) may still be below the 
resolution of the $\mu$SR technique ($\sim 10^{-2}$\,mT). 
Hence, the future use of other techniques, e.g., based on the optical 
Kerr effect~\cite{Kallin2016}, another very sensitive probe of 
spontaneous fields in unconventional superconductors, remains crucial.  
Due to their rich crystal structures, covering both centro- and 
noncentrosymmetric cases, and the pervasive presence of superconductivity 
at low temperatures, the nonmagnetic Re-based materials are 
the ideal choice for investigating the origin of TRS breaking.  
Here, we reviewed different cases of Re-containing superconductors, ranging 
from elementary rhenium, to Re$T$ ($T$ = 3$d$-5$d$ early transition metals), 
to the dilute-Re case of ReBe$_{22}$, all of which were 
investigated through both macroscopic and microscopic techniques.
By a comparative study of Re$T$ with different $T$ metals mostly using $\mu$SR technique, 
we could demonstrate the secondary role played by SOC and why the 
spin-singlet pairing is dominant in Re$T$ superconductors. 
This, however, brings up 
the question of reconciling the occurrence of TRS breaking with a 
fully-gapped SC state (spin-singlet pairing).
A possible solution to this is offered by the so-called INT model, 
which requires an antisymmetric pairing function involving the orbital degree of freedom, 
making it insensitive to the presence (or lack) of inversion symmetry 
and SOC. Overall, the reported results suggest that the rhenium 
presence and its amount are two key factors for the appearance and the 
extent of TRS breaking in the Re-based superconductors. 
These key observations, albeit important, demand new experimental 
and theoretical investigations to further generalize them.
	
To date, as nearly all current
studies have focused exclusively on 
$\alpha$-Mn-type Re$T$ superconductors (except for the Re-Mo series
considered here), the superconducting properties of most other Re$T$ 
alloys remain basically unexplored. 
Hence, the synthesis and characterization of non-$\alpha$-Mn-type Re$T$ 
alloys, including the study of their electrical, magnetic, and 
thermodynamic properties, is of clear interest.
Similarly, systematic $\mu$SR measurements, crucial for detecting 
the presence of TRS breaking in Re-based superconductors, 
are in high demand. For instance, although both Re-Zr and Re-Nb alloys 
exhibit rich crystal structures and superconducting phase diagrams, 
only their $\alpha$-Mn-type phase has been explored. 
In addition, most of the original measurements were performed only 
on polycrystalline samples. Hence, the synthesis of single crystals 
will be essential in the comprehensive search for possible superconducting nodes and, 
thus, for mixed singlet-triplet pairing. 
Finally, it would be of interest to extend the $\mu$SR studies 
on  elementary rhenium from the bulk- to its thin-film form, 
where inversion symmetry is artificially broken. By checking if the 
TRS breaking is maintained or not, will help us to further clarify the rhenium conundrum.

\section*{Conflict of Interest Statement}
The authors declare that the research was conducted in the absence of any commercial or financial relationships that could be construed as a potential conflict of interest.

\section*{Author Contributions}
The manuscript was prepared and written by both the authors. 

\section*{Funding}
This work was supported by start funding from 
East-China Normal University (ECNU), the Swiss National Science Foundation 
(Grant No.\ 200021-169455) and the Sino-Swiss Science and
Technology Cooperation (Grant No.\ IZLCZ2-170075).

\section*{Acknowledgments}
We thank M. Shi\ for the fruitful discussion. We thank M.\ Medarde for the assistance during the electrical 
resistivity and magnetization measurements, and D.\ J.\ Gawryluk 
and E.\ Pomjakushina for synthesizing the materials. We acknowledge 
the allocation of beam time at the Swiss muon source (S$\mu$S) 
(DOLLY, GPS, and LTF spectrometers).

\vspace{8pt}
\emph{Note} This is the version of article after peer review but without editing, as submitted to \emph{Frontiers in Physics}. Frontiers is not responsible for any errors or omissions in this version of the manuscript. 
The paper is available online at \url{10.3389/fphy.2021.651163}.

\bibliography{Re_review}

\end{document}